# All-electrical driving and probing of dressed states in a single spin

**Authors:** Hong T. Bui[1,2,†], Christoph Wolf[1,3,†], Yu Wang[1,3], Masahiro Haze[1,4], Arzhang Ardavan[5], Andreas J. Heinrich[1,2]*, Soo-hyon Phark[1,3]*

**Affiliations:**

[1]Center for Quantum Nanoscience, Institute for Basic Science (IBS), Seoul 03760, Korea

[2]Department of Physics, Ewha Womans University, Seoul 03760, Korea

[3]Ewha Womans University, Seoul 03760, Korea

[4]The Institute for Solid State Physics, University of Tokyo, Kashiwa 277-8581, Japan

[5]CAESR, Clarendon Laboratory, Department of Physics, University of Oxford, Oxford OX1 3PU, UK

*Corresponding authors. Email: heinrich.andreas@qns.science, phark@qns.science

†These authors contributed equally to this work.

**Abstract:**

The sub-nanometer distance between tip and sample in a scanning tunneling microscope (STM) enables the application of very large electric fields with a strength as high as ~ 1 GV/m. This has allowed for efficient electrical driving of Rabi oscillations of a single spin on a surface at a moderate radio-frequency (RF) voltage of the order of tens of millivolts. Here, we demonstrate the creation of dressed states of a single electron spin localized in the STM tunnel junction by using resonant RF driving voltages. The read-out of these dressed states was achieved all-electrical by a weakly coupled probe spin. Our work highlights the strength of the atomic-scale geometry inherent to the STM that facilitates creation and control of dressed states, which are promising for a design of atomically well-defined single spin quantum devices on surfaces.

**Introduction:**

The creation of long-lived quantum states lies at the heart of understanding quantum-coherent phenomena and their application to practical problems such as quantum computing and quantum sensing [1]. The interaction of quantum states with the environment is critical for observing the quantum state, but also inevitably leads to the uncontrollable collapse of the wave function resulting in the loss of its quantum information in a process called decoherence [2]. In such open quantum systems, various strategies have been employed to improve the coherence of their quantum state. A particularly intriguing strategy is the creation of decoherence-free sub-spaces, which interact with the environment slowly compared to the dynamics of the quantum states, leading to a suppression of decoherence.

One feasible approach to such quantum states with long coherence times is to create so-called dressed states, which result from the coupling between the quantum spins and a coherent driving field, also known as the AC-Stark effect [3,4]. The consequence of a coupling between a resonant driving field and an atomic spin is vacuum Rabi oscillations as long as the coupling between spin and field is larger than the incoherent dynamics of the system, which contribute to the decay of Rabi oscillations [5]. In this strongly driven case, the spin eigenstates of the system are dressed with the resonant field. The dressed states can be probed by using an optical transition to a third level, where they appear as Autler-Townes doublets [3,6] or Mollow Triplets [7,8] in various two-level systems, such as cold atomic gasses [9-11], single molecules [12,13], quantum dots [14,15], and defects in semiconductor [4]. The driving field can be implemented over a wide range of energy, from lasers [16] to microwave [4,17].

In this work, we demonstrate the creation and read-out of dressed states of single spins in a scanning tunneling microscope (STM) [18,19]. The STM tunnel junction with a characteristic gap of about one nanometer easily generates large electric fields on the order of ~ 1 GV/m, in contrast to centimeter-scale geometries common for similar experiments on bulk samples [20,21]. Therefore, a spin confined in the STM junction can fall into the strong driving regime when driven by an oscillating electric field. Here we utilized a tailored atomic scale nanostructure of two weakly coupled electron spins (Fig. 1a), in which multi-spin control was achieved through a spin in the STM tunnel junction using the double resonance technique [22,23]. Both spins were subjected to the applied radio frequency (RF) fields and selectively driven by selectively tuning the RF frequency into resonance [24-28]. Resonant driving of one spin resulted in dressing of its states. Instead of using an optical transition to a third level, resonant transitions of the other spin were then used to probe the dressed states of the first spin in our experimental scheme, enabling both creation and read-out of the dressed states to be performed all-electrical with two RF voltages of moderate amplitude. Our scheme yielded Autler-Townes doublets as well as Mollow triplets in the double resonance spectra and, in addition, provided a direct way to measure the Rabi rates of both spins in a continuous-wave experiment.

**Experiment:**

We performed experiments on a designed nanostructure composed of two Ti atoms (each having electron spin of S = 1/2), hereafter referred to as Ti-1 and Ti-2, forming a pair of coupled electron spins. These spins were placed on a bilayer of MgO grown on a single crystalline Ag substrate (Fig. 1a) [29, 30]. The electron spin resonance (ESR) spectra of Ti-1 (Fig. 1b) and Ti-2 (Fig. 1c) were obtained by using continuous-wave single- and dual-frequency ESR, respectively[22]. From the spectra, we obtained the respective resonance frequencies of the coupled spins as well as the strength of their exchange interaction $J$. The eigenstates of this system can be well described by Zeeman product states $|S_z(\text{Ti-1})\ S_z(\text{Ti-2})\rangle$, yielding four possible transitions (with frequencies $f_1$ to $f_4$), which are addressable by tuning the frequency of the RF driving voltages, as depicted in Fig. 1d.

Driving any of these four transitions close to their resonance frequency leads to dressing of the spin corresponding to the transition, which results in new eigenstates formed by the superposition of two relevant Zeeman product states of the undriven system. These new eigenstates, called dressed states, are labeled as $|D\rangle$ in Fig. 1d. The energy splitting $\Delta f$ between a pair of dressed states corresponds to the Rabi rate $\Omega$ [31, 32] determined by the coupling between the two-level system and the driving field [3]. Our specific design of the double resonance measurement schemes [22] for a coupled spin system allows such dressed states to be spectroscopically probed by another transition to a state which does not undergo state dressing. To simultaneously dress and probe the spin states, we used two RF signal sources where we fixed the frequency of one RF voltage to dress one spin and swept the other across the resonances of the second spin to spectroscopically probe the dressed states as depicted schematically in Fig. 1d. All the spectra shown in Figs. 1–4 were obtained with the tip positioned on Ti-1.

**Driving and probing of dressed states:**

We begin by dressing the Ti-1 spin and probing its splitting using a transition of Ti-2. In Fig. 2a, we show double resonance spectra obtained with varying RF voltage $V_{RF1}$ with its frequency ($f_{RF1}$) fixed at a resonance of Ti-1 ($f_1$) and sweeping the frequency of the other RF voltage ($f_{RF2}$) across a resonance of Ti-2, $f_3$ (see Fig. 2b). Note that each spectrum was measured with the tip positioned on Ti-1, i.e., the spin-polarized tunneling current measures the ESR signal of the Ti-1 spin, which remained in resonance by the given $V_{RF1}$ at $f_1$ during the frequency sweep of $V_{RF2}$. Our double resonance scheme is designed to read out zero intensity for the ESR signal of Ti-1 when Ti-2 is off-resonance, such that any non-zero intensity of a double resonance spectrum reflects a variation of the Ti-1 spin's resonance intensity in the frequency sweep of $V_{RF2}$ [22]. Therefore, the peaks shown in Fig. 2a indicate a net reduction of the ESR signal of Ti-1 due to spin population transfer from $|00\rangle$ to $|01\rangle$, stemming from the resonance of the Ti-2 spin at $f_{RF2} = f_3$.

The formation of dressed states manifested as splitting of the double resonance peak, which increased monotonically for an increasing RF voltage $V_{RF1}$. The dressing of the Ti-1 spin, followed by the probing using Ti-2 transitions, is illustrated in Fig. 2b. The two states $|00\rangle$ and $|10\rangle$ relevant to the transition $f_1$ are dressed, resulting in two new eigenstates, $|D\rangle_{00\pm}$ and $|D\rangle_{10\pm}$, respectively. On driving the Ti-2 spin across $f_3$, the two dressed states stemming from $|00\rangle$ were probed as two resonant transitions $\alpha$ and $\beta$ as depicted. We emphasize here that the Ti-2 spin probes the dressed states of the Ti-1 spin. The splitting $\Delta f$ was visible when it was larger than the linewidths of the double resonance peaks and showed a linear dependence on $V_{RF1}$. From the linear fit in Fig. 2c we extracted the Rabi rate of the Ti-1 spin, $\Omega^{(1)}/(2\pi V_{RF1}) = 0.160 \pm 0.015$ MHz/mV.

Next, we exchanged the roles of the two spins by dressing Ti-2 and probing with Ti-1. To achieve this in our double resonance experiment, we fixed $f_{RF2}$ at the resonance $f_3$ of Ti-2 and swept $f_{RF1}$ over the frequency range of the Ti-1 transitions. The resulting spectra are shown in Fig. 3a. When increasing the driving RF voltage ($V_{RF2}$), we again observed splitting of the double resonance peaks. In contrast to the case in Fig. 2, the dressing of the Ti-2 spin at frequency $f_3$ splits the states, $|00\rangle$ and $|01\rangle$, into four dressed states, $|D\rangle_{00\pm}$ and $|D\rangle_{01\pm}$, respectively, as illustrated in Fig. 3b. Then, the dressed states were probed by the resonances $f_1$ and $f_2$ of Ti-1. The peak splitting $\Delta f$ again showed a linear dependence on $V_{RF2}$, resulting in a Rabi rate of the Ti-2 spin, $\Omega^{(2)}/(2\pi V_{RF2}) = 0.220 \pm 0.02$ MHz/mV (Fig. 3c). Here, the Autler-Townes doublets in the double resonance spectroscopy allowed a direct measure of the strength of the remote driving of a spin outside of the tunnel junction as reported in our previous work [22, 23] Note that the Rabi rate of the Ti-2 spin is comparable to that of the Ti-1 spin in the tunnel junction, demonstrating that our tailored quantum spin structure provides an effective way to remotely control the spins outside the STM tunnel junction.

We extended our double resonance scheme to observe the Mollow triplet [7, 8] from the dressed states of the Ti-2 spin. Here, we first measured a double resonance spectrum using the same scheme in Fig. 3b by driving only the transition $f_3$. As a result, the spectrum showed four peaks composed of two groups, where each group is the Autler-Townes doublet measured by the probing transition corresponding to either $f_1$ or $f_2$ (Fig. 4a). Then, we simultaneously drove both resonances ($f_3$ and $f_4$) of Ti-2 and measured the spectrum across the spin resonances of Ti-1. The resulting spectrum in Fig. 4b clearly shows six peaks composed of two groups, three peaks per group across each resonance of the Ti-1 spin. The simultaneous driving of both transitions $f_3$ and $f_4$ dressed all four states $|00\rangle, |01\rangle, |10\rangle, |11\rangle$, leading to four transitions ($\alpha', \beta', \gamma', \delta'$) available when probed across each resonance of Ti-1, as illustrated in Fig. 4c. The transitions $\beta'$ and $\gamma'$ have the same frequency as the un-dressed states, whilst two others, $\alpha'$ and $\delta'$, are shifted by $-\Omega$ and $+\Omega$, respectively. This resulted in three peaks in each group, where the intensity of the central one is twice as large as those of the two side peaks, as seen clearly in the spectrum.

**Simulations:**

To understand the spin dynamics of this work, we performed open quantum system simulations of two exchange-coupled spins using the Lindblad formalism with collapse operators to account for finite lifetime and coherence times of the spins (see also Supplementary 4) [33]. All the spectra of the Autler-Townes doublet were well reproduced by implementing double resonance spectroscopy schemes into the simulations and calculating the spin-polarized current using steady-state populations of the four quantum states (gray curves in Figs. 2 and 3). Our simulation using three RF driving voltages produced spectra also in quantitative agreement with the Mollow triplets in Fig. 4b (blue overlay spectrum at $V_{RF2}$ = 50 mV) and estimated its driving-power dependence (Fig. 4d). Deviations in the relative peak heights are sensitive to detuning of the driving frequencies, which might stem from slight deviations of an experimental factor such as the fluctuation of the external magnetic field (see Fig. S5). We note that given our experimental parameters the average number $\langle n \rangle$ of RF photons is large, and thus the resulting dressed states can be described using a semi-classical photon field [34-36].

Further insights into the state-dressing can be found by analyzing the time evolution of the density matrix elements. Our experimental scheme, using two coupled spins, differs from the conventional schemes by the all-electrical driving and readout of the dressed states [37]. This, however, turned out not to significantly influence the basic principle of state dressing, which was shown by tracing out the probe spin and analyzing the reduced density matrix only for the dressed spin $\rho^{(2)} = \text{Tr}_1[\rho]$. Here, the indices 1 and 2 refer to probe and dressed spin, respectively. After tracing out the sensor spin, the resulting density matrix of the dressed spin, composed of two diagonal elements, $\rho_{00}$ and $\rho_{11}$, corresponding to the populations of the $|0\rangle$ and $|1\rangle$ states, respectively, and two off-diagonal elements, $\rho_{01}$ and $\rho_{10}$, correspond to the coherences of the dressed spin (see Supplementary 4 for the details).

We compare in Figs. 5a and 5b two simulated density matrices of the dressed spin driven at two driving RF amplitudes $V_{RF2}$ = 20 and 140 mV, respectively, where the probe spin was weakly driven in all simulations at a fixed $V_{RF1}$ = 30 mV, which is the same as in the experiments. A weakly driven system ($V_{RF2}$ = 20 mV; Fig. 5a) showed that its populations, $\rho_{00}$ and $\rho_{11}$, just decay slowly towards the steady states, in which the population distribution was determined by the driving amplitude. In contrast, a strongly driven system ($V_{RF2}$ = 140 mV; Fig. 5b) showed clear Rabi oscillations of the populations in the transient (0 < t < 50 ns), indicating that here the driving power fulfilled the condition for the strong driving regime, where the spin-field coupling was larger than the decoherence dynamics of the system. The populations also decayed in the long-time limit, however, saturating to near-equal values for $|0\rangle$ and $|1\rangle$ states, which are close to the limit in the continuous wave ESR. All our simulations were implemented in the lab frame, leading to additional fast oscillations as shown in the insets in Figs. 5a and 5b.

The time evolution of the off-diagonal elements incorporates the coherence of the system, i.e. the correlations between the states $|0\rangle$ and $|1\rangle$, as well as the Larmor precession. When the spin is coupled with a driving field, in addition, the off-diagonal elements become modulated by the Rabi coupling since the system eigenstates transform into the dressed states $|D\rangle_{\pm}$, which are superpositions of the unperturbed eigenstates, $|0\rangle$ and $|1\rangle$. The coherence element $\rho_{01}$ at $V_{RF2}$ = 20 mV also slowly decayed towards a steady state, with dense oscillations due to the very fast Larmor precession at the transition frequency. To show the frequency spectrum of the signal we took the Fourier transform of the time evolution of $\rho_{01}$ (Fig. 5c). It shows a strong central peak at the transition frequency of spin 2 ($f_3$) and two weak satellites. The separation between these peaks is too small, and as a consequence, the state dressing in a weakly driven system was not observed in the experiments. In contrast, the Fourier transform of $\rho_{01}$ at $V_{RF2}$ = 140 mV (Fig. 5d) shows two clearly separated satellite peaks, with the separation being linearly dependent on the driving amplitude (see Fig. 5e). We note that in the experiment the splitting between two peaks ($\alpha, \beta$) is given by $\Omega$ since the transition is probed by a third level which is not dressed (see also Fig. 1d). Similarly, the two side band peaks observed in the density matrix are split by $\pm\Omega$ from the center frequency, as illustrated in Fig. 5f. The simulations shown here strongly suggest that our designed, weakly coupled two-spin system, combined with the double resonance scheme, can in good approximation be treated as a single dressed two-level system when one of the two spins is strongly driven (see also Fig. S5).

**Conclusion & Outlook:**

We demonstrated that dressing spin states of an individual atom with resonant RF fields is easily achievable in an ESR-STM at moderate RF voltages owing to the inherently strong electric field in the sub-nm geometry of the STM junction. By utilizing two coupled spins and continuous wave RF excitations at two or three frequencies simultaneously, we measured Autler-Townes doublets and Mollow triplets in the double resonance spectra. From the splitting of the doublet and triplet, we were able to determine the Rabi rate of each spin independently. We anticipate that the techniques shown here are critical for creating decoherence-free subspaces with long-lived quantum states on surfaces which might become crucial for moving this platform towards real implementation of quantum algorithms.

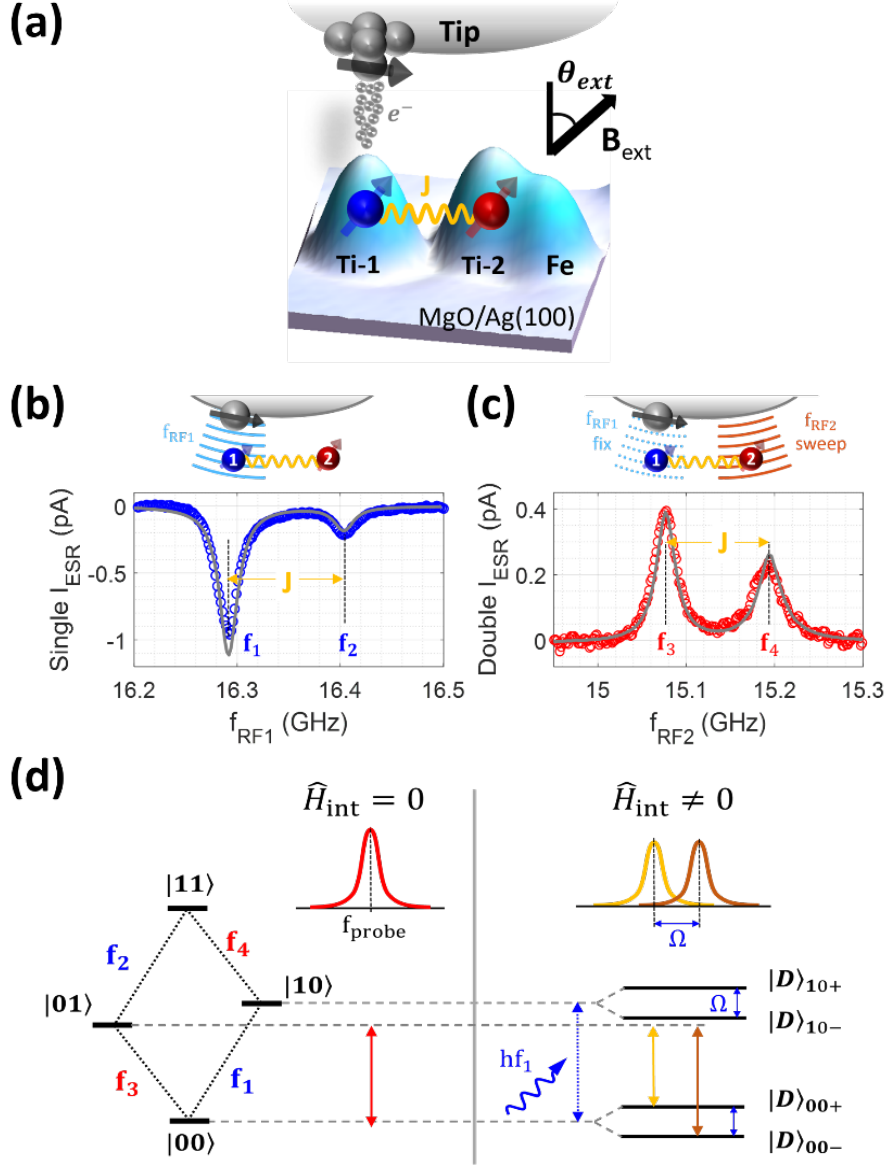

**Figure 1. Dressing of a single spin in the STM tunnel junction by coupling with an oscillating field. (a)** STM topographic image of an engineered nanostructure composed of two coupled spins: Ti-1 (in the tunnel junction), Ti-2 (outside the tunneling current), and Fe adatom. The Ti-1 to Ti-2 spacing is 1.22 nm, and Ti-2 to Fe is 0.59 nm ($I_{DC}$ = 10 pA, $V_{DC}$ = 100 mV) [22]. **(b)** and **(c),** Single and double resonance ESR spectra obtained from the structure in (a). In (b) the spectrum reveals two transitions of Ti-1 (labeled as $f_1$ and $f_2$), whilst in (c) the spectrum shows two transitions ($f_3$, $f_4$) of Ti-2 ($I_{DC}$ = 20 pA, $V_{DC}$ = 50 mV, $B_{ext}$ = 670 mT, $\theta_{ext}$ = 60°, $V_{RF1}$ = 30 mV; $V_{RF2}$ = 40 mV for (c)). The grey curves are simulated spectra. The splitting of the peaks in each case corresponds to the exchange coupling energy $J$ between the two Ti spins. **(d)** Energy diagrams of two weakly coupled spins with four spin states labelled according to $|S_z(\text{Ti-1}), S_z(\text{Ti-2})\rangle$ and ESR transitions $f_1, f_2, f_3$, and $f_4$ corresponding to the four peaks shown in (b) and (c). Dressed-state picture of a quantum two level system, here for the first spin (Ti-1) for the cases of spin-field coupling OFF ($\widehat{H}_{\text{int}} = 0$) and ON ($\widehat{H}_{\text{int}} \neq 0$). The two relevant states ($|00\rangle, |10\rangle$) are dressed ($|D\rangle_{10\pm}, |D\rangle_{00\pm}$) when the coupling is switched ON. The transition $f_3$ is used to probe the dressed states.

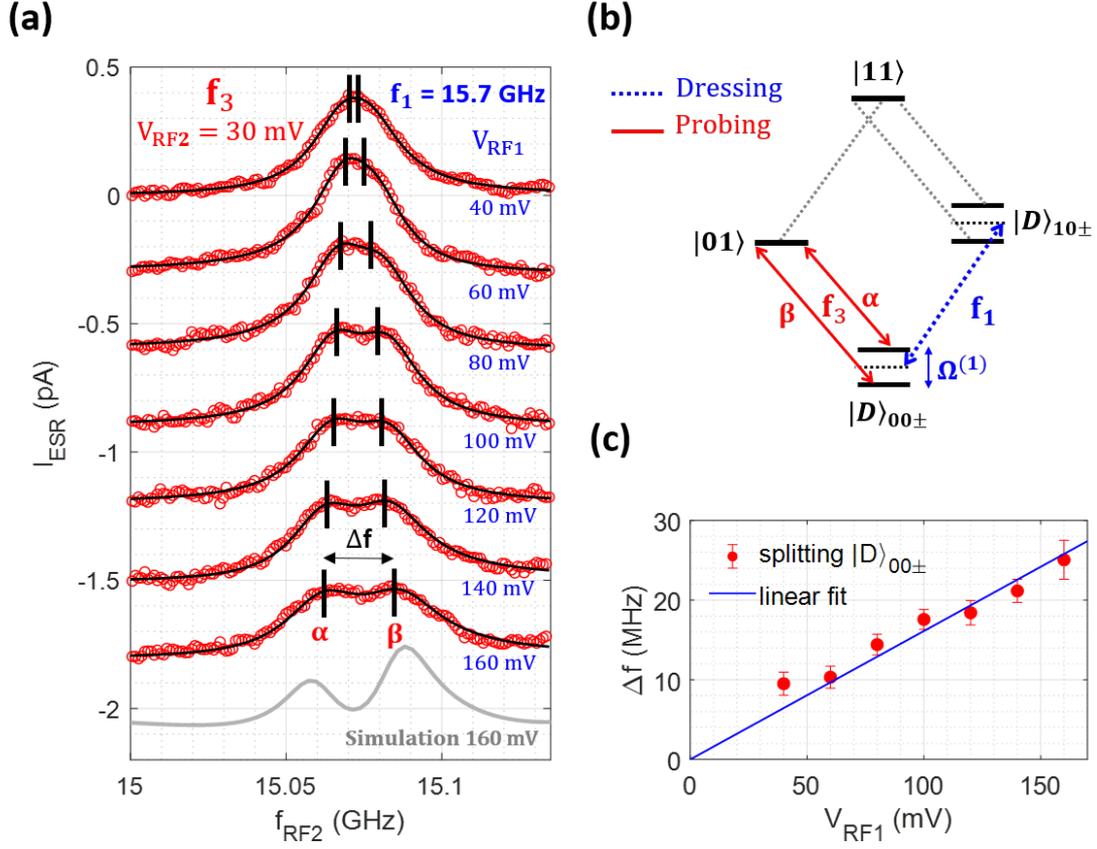

**Figure 2. Double resonance spectroscopy on dressed states of Ti-1.** (**a**) Double resonance spectra measured with RF2 frequency swept across an ESR transition ($f_3$) of Ti-2 and with RF1 frequency fixed at a resonance of Ti-1 ($I_{DC}$ = 20 pA, $V_{DC}$ = 50 mV, $V_{RF2}$ = 30 mV, $T$ = 0.4 K, $B_{ext}$ = 670 mT, $\theta_{ext}$ = 60°). An increasing driving amplitude of Ti-1 ($V_{RF1}$) led to an increased peak splitting (see $\alpha$, $\beta$). The curves were fitted using two-peak Lorentzian function (solid black curves). The gray curve is a simulated spectrum using the experimental peak splitting as Rabi rate and spin relaxation times of $T_1^{(1)}$ = 8 ns, $T_1^{(2)}$ = 150 ns for the first and second spins, respectively. (**b**) A schematic energy level diagram illustrating the dressing of Ti-1 ($f_1$) and probing using Ti-2 ($f_3$). (**c**) Dependence of the splitting $\Delta f = \alpha - \beta$ on the driving amplitude $V_{RF1}$ as obtained from the fitting of the spectra in (a). The solid line is a linear fit, giving the Rabi rate of the transition ($f_1$), $\Omega^{(1)}/V_{RF1}$ = 0.16 ± 0.015 MHz/mV.

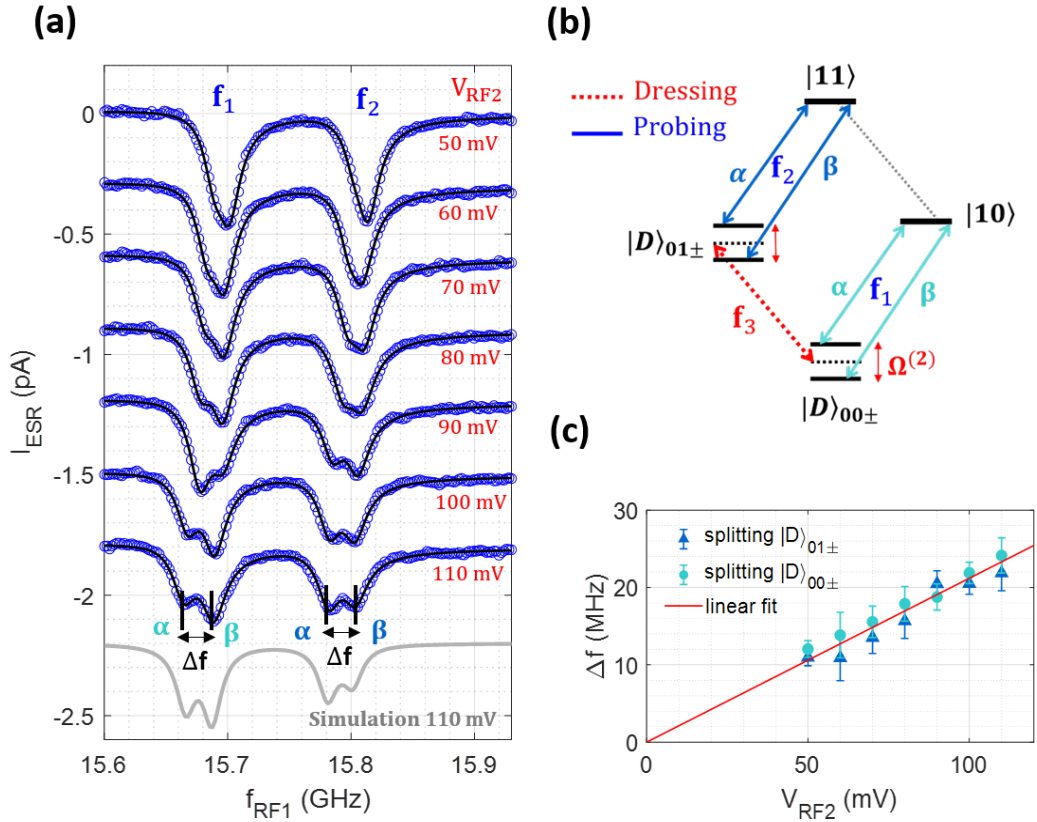

**Figure 3. Double resonance spectroscopy on dressed states of Ti-2.** (**a**) Double-resonance spectra measured with RF1 frequency swept across the ESR transitions ($f_1$, $f_2$) of Ti-1 and with RF2 frequency fixed at a resonance ($f_3$) of Ti-2 ($I_{DC}$ = 20 pA, $V_{DC}$ = 50 mV, $V_{RF1}$ = 30 mV, $T$ = 0.4 K, $B_{ext}$ = 660 mT, $\theta_{ext}$ = 70°; also see Fig. S4 for the four transition frequencies). An increasing driving amplitude of Ti-2 ($V_{RF2}$) led to an increased peak splitting (see $\alpha$, $\beta$). The gray solid curve represents a simulated spectrum using the peak splitting as Rabi rate and spin relaxation times of $T_1^{(1)}$ = 8 ns, $T_1^{(2)}$ = 150 ns for the first and second spins, respectively. (**b**) A schematic energy level diagram illustrating the dressing of Ti-2 ($f_3$) and probing using Ti-1 ($f_1$, $f_2$). (**c**) Dependence of the splitting $\Delta f = \alpha - \beta$ on the driving amplitude $V_{RF1}$ obtained from the fitting of the spectra in (a). The solid line is a linear fit, giving the Rabi rate of the transition ($f_3$), $\Omega^{(2)}/V_{RF2}$ = 0.22 ± 0.012 MHz/mV.

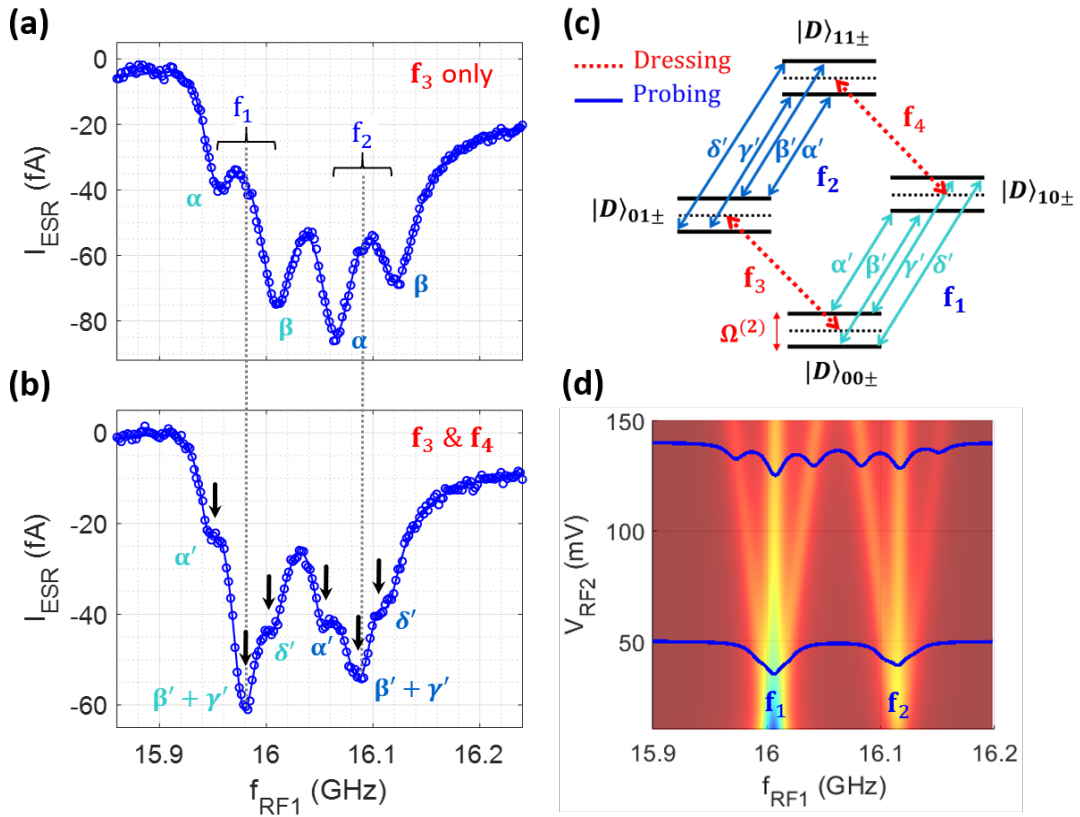

**Figure 4. Mollow triplet from dressed states of Ti-2.** (**a**) Double resonance spectra showing Autler-Townes doublets from the dressed states of Ti-2 at its transition $f_3$ with $V_{RF2}$ = 120 mV and $V_{RF1}$ = 20 mV. (**b**) Triple resonance spectra measured by applying $V_{RF2}$ simultaneously on two transitions of Ti-2 ($f_3$ and $f_4$) with equal amplitudes, $V_{RF2}(f_3) = V_{RF2}(f_4) = 60$ mV, and on probing transitions ($f_1$ and $f_2$) $V_{RF1}$ = 20 mV. Mollow triplets were observed (three peaks per probing transition), resulting from the 8 possible transitions across the probing transition frequencies, $f_1$ and $f_2$, as depicted in (c) ($I_{DC} = 20$ pA, $V_{DC} = 50$ mV, $T = 0.4$ K, $B_{ext} = 640$ mT, $\theta_{ext} = 65°$; also see Fig. S4 for the four transition frequencies). The tip used in this measurement was different from the tip for Figs. 2 and 3. (**c**) Schematic of dressed states of Ti-2 under simultaneous driving of its two resonances $f_3$ and $f_4$. (**d**) Simulated Mollow triplet spectra vs. $V_{RF2}$ amplitude, showing an increased splitting for an increasing $V_{RF2}$, with two representative spectra overlayed in blue.

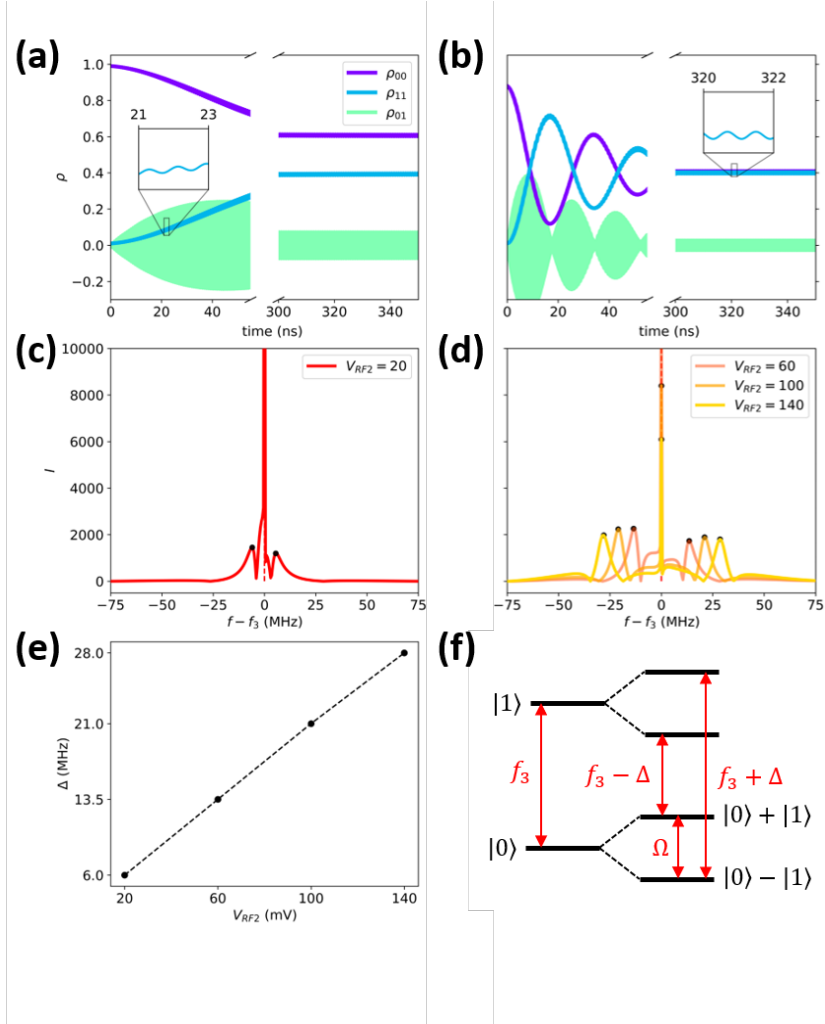

**Figure 5. Dynamics of the dressed spin.** Difference between the time-evolution of a weakly and strongly driven spin obtained from the reduced density matrix of the second spin (Ti-2). (**a**) time evolution of the diagonal elements, $\rho_{00}$ and $\rho_{11}$, the populations of ground $|0\rangle$ and excited state $|1\rangle$ of a weakly driven ($V_{RF2}$ = 20 mV) spin, respectively, reveals overdamped evolution of the populations towards the steady states. (**b**) in contrast, a strongly ($V_{RF2}$ = 140 mV) driven spin shows Rabi oscillations in the transient (0 < t < 50 ns) and near-equal populations in the long-time limit. (**c**) Fourier transform of the time evolution of the coherence, $\rho_{01}$ in the weakly driven case, showing a strong peak at the transition frequency ($f_3$). (**d**) Fourier transforms of $\rho_{01}$ with an increasing driving amplitude. The satellite peaks corresponding to $f_3 \pm \Delta$ become clearly visible and shift linearly with the driving amplitude as shown in (**e**). (**f**) Energy level scheme of the dressed spin, without (left) and with state dressing (right). Note that here $\Delta = \Omega$, the same as the experimental splitting measured via a second spin.

# SUPPLEMENTARY MATERIALS

**Supplementary 1: AC Stark effect in double resonance spectra**

In this work, we demonstrated the AC Stark effect in a single spin coupled with a radio-frequency electric field in STM tunnel junction. An elegant approach to understand the AC Stark effect is the "dressed atom" theory as follows. Consider a quantum system of two atom levels, $|0\rangle$ and $|1\rangle$ ($E_{|0\rangle} < E_{|1\rangle}$), in a near-resonant electromagnetic field with photon energy $\hbar\omega$, so that $E_{|1\rangle} - E_{|0\rangle} \approx \hbar\omega$. If photons do not interact with the atom states, the total Hamiltonian is $\widehat{H}_0 = \widehat{H}_{\text{atom}} + \widehat{H}_{\text{field}}$, and the eigenstates of $\widehat{H}_0$ are $|0, n\rangle$ and $|1, n\rangle$ where $n$ labels the photon occupation. We notice that the state of the atom level $|1\rangle$ with the photon number $n - 1$ ($|1, n-1\rangle$) is almost degenerate with the state $|0, n\rangle$, and so are $|1, n\rangle$ and $|0, n+1\rangle$. We define the detuning as $\Delta = E_{|1,n-1\rangle} - E_{|0,n\rangle}$, and the two states $|0, n\rangle$ and $|1, n-1\rangle$ become fully degenerate at $\Delta = 0$.

Now if we turn on an interaction $\widehat{H}_{\text{int}}$ between atom and field, the states $|0, n\rangle$ and $|1, n-1\rangle$ become hybridized and their degeneracy lifted. Then, the eigenstates of the system can be described by a linear combination of the two nearly degenerate states of the non-interacting Hamiltonian

$$|D\rangle = C_0|0, n\rangle + C_1|1, n-1\rangle, \qquad \text{(Eq. S1)}$$

where the coefficients are obtained from the Schrodinger equation,

$$(\widehat{H}_{\text{atom}} + \widehat{H}_{\text{field}} + \widehat{H}_{\text{int}})|D\rangle = E_D|D\rangle. \qquad \text{(Eq. S2)}$$

Defining the Rabi rate as $\hbar\Omega \equiv \langle 1, n-1|\widehat{H}_{\text{int}}|0, n\rangle$, we can solve the secular equation and obtain the eigenenergies

$$E_D = (n - 1/2)\hbar\omega \pm \hbar\Omega_R/2, \quad \Omega_R = \sqrt{\Delta^2 + \Omega^2}, \qquad \text{(Eq. S3)}$$

and the corresponding eigenstates,

$$|D(n)\rangle_+ = \sin\theta\,|0, n\rangle + \cos\theta\,|1, n-1\rangle, \qquad \text{(Eq. S4)}$$
$$|D(n)\rangle_- = \cos\theta\,|0, n\rangle - \sin\theta\,|1, n-1\rangle, \qquad \text{(Eq. S5)}$$

where $\cos 2\theta \equiv \Delta/\Omega_R$ and $\sin 2\theta \equiv \Omega/\Omega_R$. At resonance ($\Delta = 0$), $\sin\theta = \cos\theta = 1/\sqrt{2}$, and the new eigenstates are split by the Rabi rate $\Omega$ due to the atom-field interaction $\widehat{H}_{\text{int}}$.

In our double resonance experiment, four spin states are involved. However, in each experiment, only one ESR transition was strongly driven. For example, only the transition $f_1$ between states $|00\rangle$ and $|10\rangle$ as shown in Fig. 2b. As a result, the ESR transition between the states $|00\rangle$ and $|01\rangle$ occurs at two distinct frequencies $f_3^{(\pm)} = f_3 \pm \Omega^{(1)}/2$ when probed by another weak electromagnetic field that was swept across the Ti-2 resonance $f_3$. The splitting allows us to determine the Rabi rate of Ti-1 as $\Omega^{(1)}/(2\pi V_{RF1}) = 0.160 \pm 0.015$ MHz/mV (Fig. 2c). The ESR peak heights at $f_3^{(\pm)}$ are sensitive to the detuning $\Delta$ due to its influence on the coefficients $\sin\theta$ and $\cos\theta$ in Eqs. S4 and S5, which is also seen in our simulations in Fig. S5. We omit the quantum numbers of the photon states for convenience in the main figures and text.

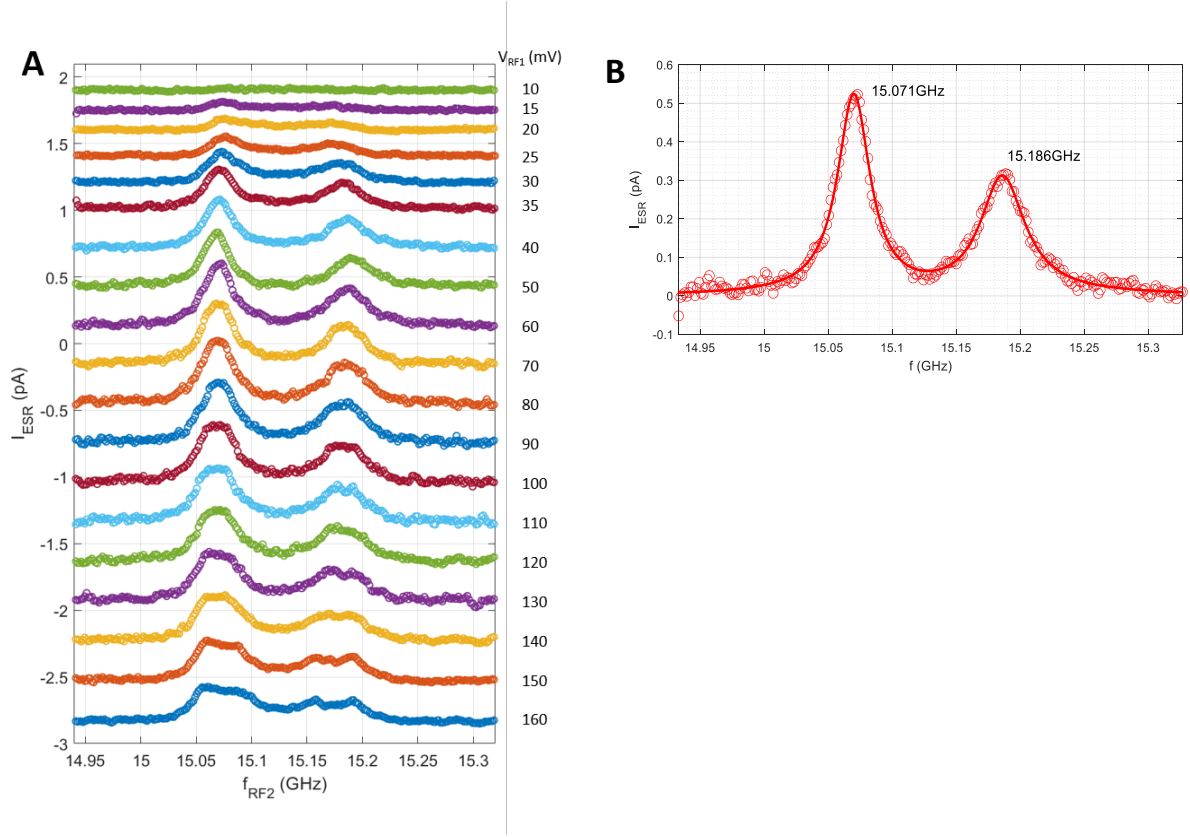

**Figure S1. Complete set of double resonance spectra across the two resonances of Ti-2, $f_3$ and $f_4$.** (A) double resonance spectra measured with $f_{RF1}$ fixed at $f_1 = 16.289$ GHz over a wide range of $V_{RF1}$ ($I_{DC} = 20$ pA, $V_{DC} = 50$ mV, $V_{RF2} = 30$ mV, $0.4 < T < 0.5$ K). (B) 2-Lorentzian curve fit to double resonance spectrum at $V_{RF1} = 60$ mV, yielding the resonance frequencies of Ti-2 spin, $f_3$ and $f_4$, in the absence of the tip's magnetic field.

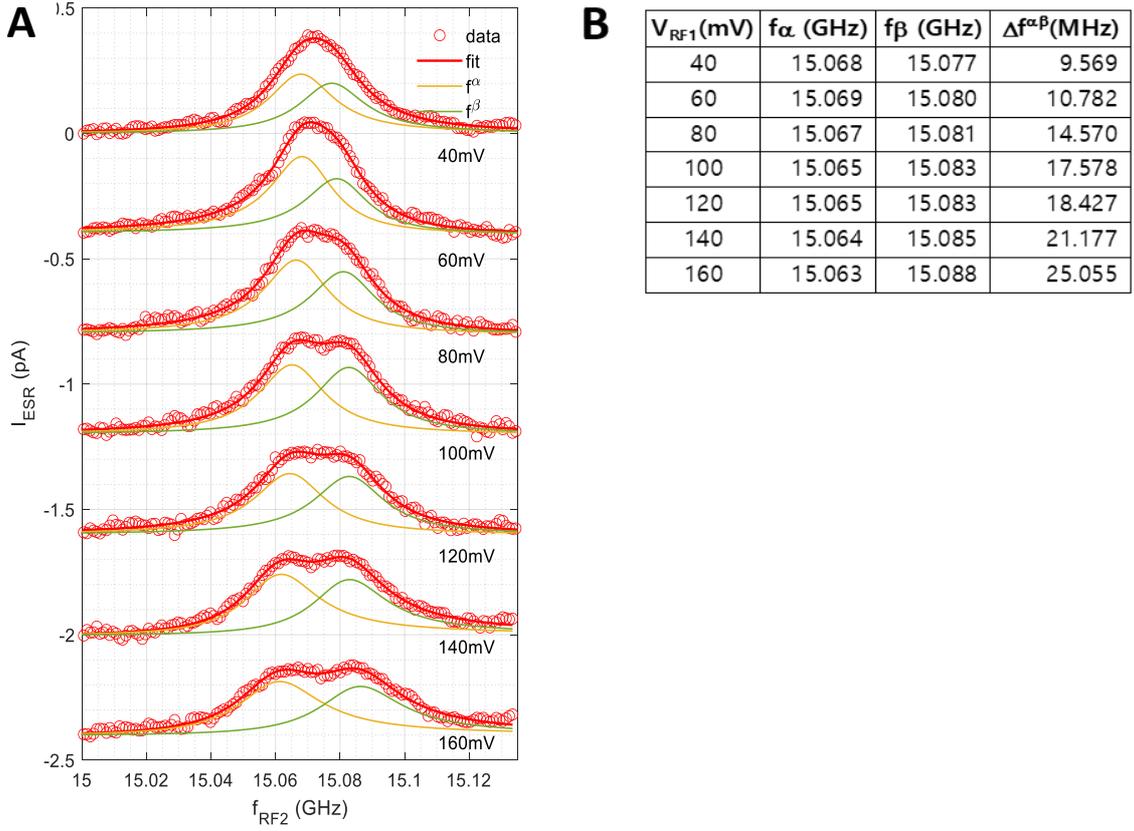

**Figure S2. Fitting of double resonance peak splitting in Fig. 2a.** (A) 2-Lorentzian curve fits to double resonance spectra in Fig. 2a. For each curve, the two peaks were fitted with one common peak width and two independent peak heights. (B) Peak frequencies ($f_3^\alpha$, $f_3^\beta$) and splitting $\Delta f^{\alpha\beta} = f_3^\beta - f_3^\alpha$, extracted from the fits in A.

**Supplementary 2: Autler-Townes doublets from dressing Ti-2 at $B_{\text{ext}}$ used for data in Fig. 2**

We also observed Autler-Townes doublets by dressing Ti-2 (see Fig. S3) at $B_{\text{ext}}$ used for the data in Fig. 2. Compared to the data shown in Fig. 3, a slightly larger out-of-plane component of $B_{\text{ext}}$ here resulted in a correspondingly larger Rabi rate of Ti-2, $\Omega^{(2)}/(2\pi V_{\text{RF1}}) = 0.260$ MHz/mV, as shown in Fig. S3B. This indicates that the interaction between Ti-2 and Fe increases with the out-of-plane component of $B_{\text{ext}}$, leading to a stronger driving of the Ti-2 spin and bigger AC Stark splitting.

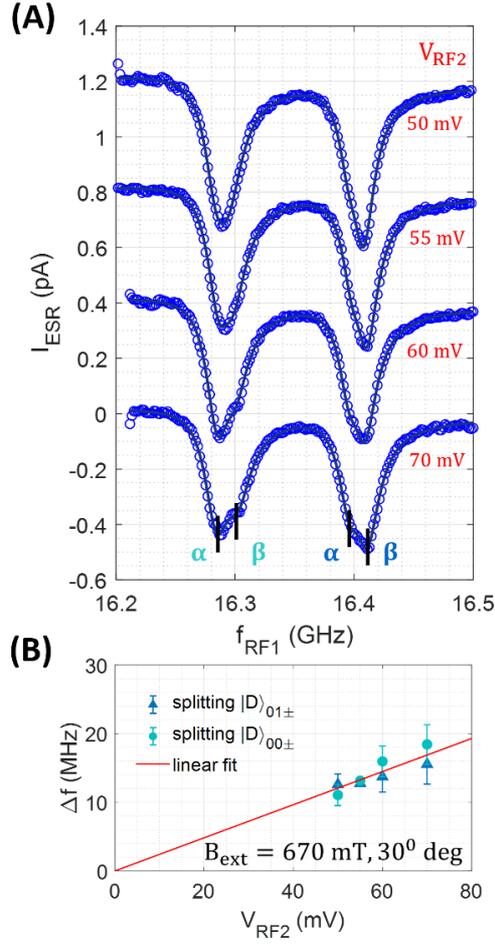

**Figure S3. Autler-Townes doublets probed using the transitions of Ti-1.** (A) Ti-1 ESR spectra with $f_{RF2}$ fixed at $f_3 = 15.071$ GHz by varying $V_{RF2}$ from 50 to 70 mV, measured with the same scheme used in Fig. 3a. Solid curves overlayed on the data: 4-Lorentzian curve fits to each spectrum. For each curve, the two subpeaks were fitted with one common peak width and two independent peak heights ($I_{DC} = 20$ pA, $V_{DC} = 50$ mV, $V_{RF1} = 30$ mV, $0.4 < T < 0.5$ K). (B) $V_{RF2}$-dependence of the splitting extracted from the fits of the spectra in A. Solid line: linear fit to the splitting.

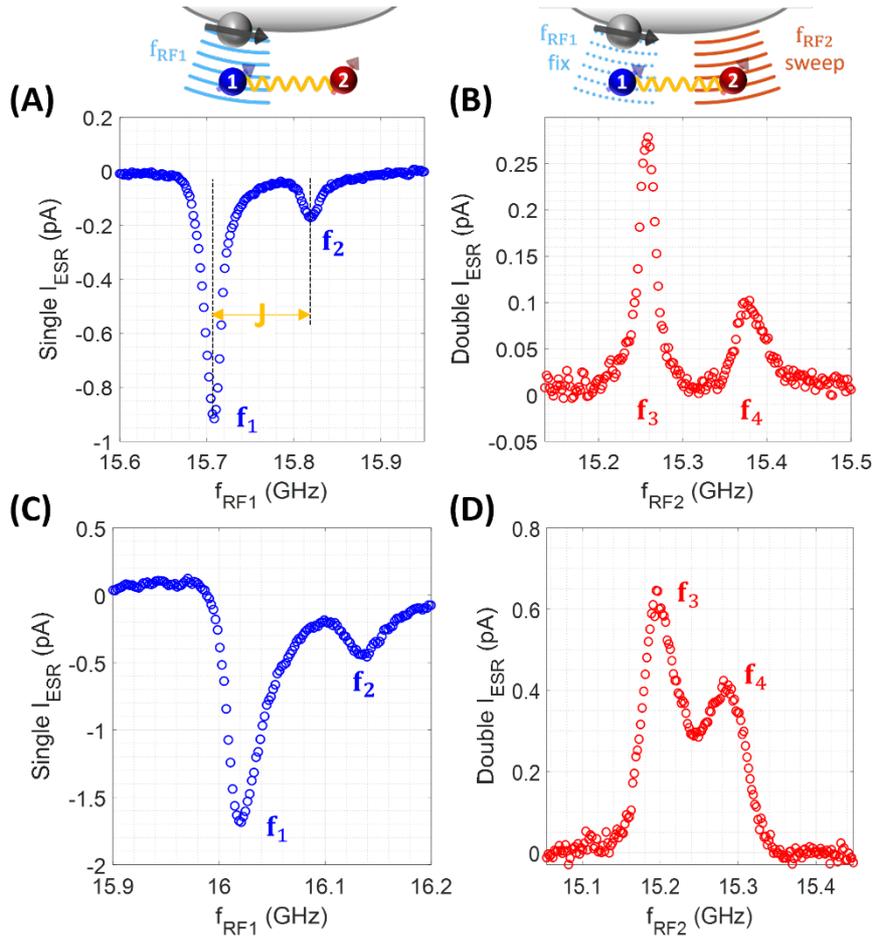

**Figure S4. Representative ESR spectra for the data in Figs. 3 and 4.** (**A** and **B**) Ti-1 single resonance and Ti-2 double resonance ESR spectra with the same tip and at the same measurement conditions used in Fig. 3. (**C** and **D**) Ti-1 single resonance and Ti-2 double resonance ESR spectra with the same tip and at the same measurement conditions used in Fig. 4. All spectra were measured at $I_{DC} = 20$ pA, $V_{DC} = 50$ mV, $V_{RF1} = 30$ mV, $V_{RF2} = 30$ mV, $T = 0.4$ K.

## Supplementary 3: Detuning effect

A small detuning ($\delta \equiv f_{RF} - f_{res} < 10$ MHz) can change the relative peak heights of the Autler-Townes doublet as shown in Fig. S5. Such detuning can occur in the experiment due to imperfect frequency tuning in the RF generator or external magnetic field fluctuations.

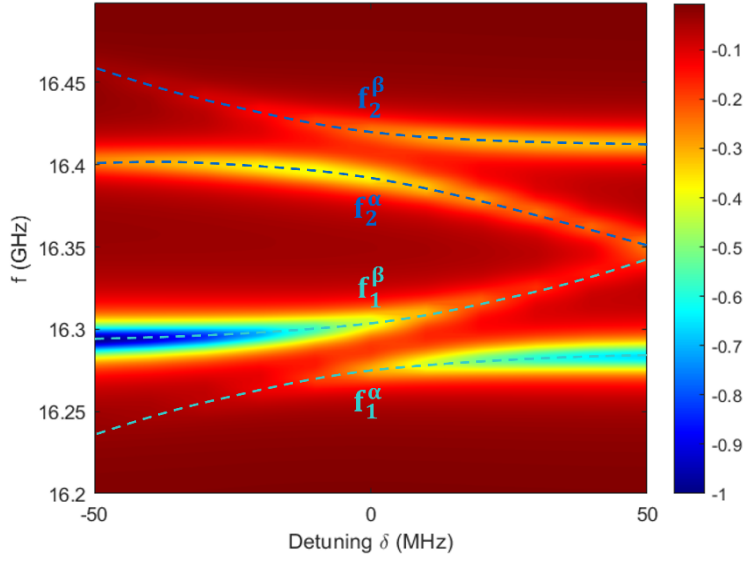

**Figure S5. Effect of detuning of the ESR driving on simulated spectra.** Simulated Ti-1 double resonance spectra for varying detuning $\delta$ of $f_{RF2}$ for $V_{RF2} = 70$ mV, $T_1^{(1)} = 8$ ns, $T_1^{(2)} = 150$ ns at the magnetic field used in Fig. 1. Vertical lines correspond to the sub-peaks of the spectrum.

## Supplementary 4: Dressed density matrix

To obtain the frequency decomposition of the density matrix we first calculated the reduced density matrix of the dressed spin by tracing out the first spin, i.e., $\rho^{(2)} = \text{Tr}_1[\rho]$, where the superscript denotes the spin. The resulting $2 \times 2$ density matrix of the dressed spin reads the Eq. (1)

$$\rho^{(2)} = \begin{bmatrix} \rho_{00} & \rho_{01} \\ \rho_{10} & \rho_{11} \end{bmatrix}, \qquad (1)$$

which is composed of 4 components representing the spin dynamics only for the second spin (Ti-2), two for its populations $\rho_{00} = p_0^{(2)}$, $\rho_{11} = p_1^{(2)}$ on the diagonals and the other two for its coherences $\rho_{01}, \rho_{10}$ on the off-diagonals. The spectra in Figs. 5c and 5d were obtained by numerical Fourier transformation of the off-diagonal element $\rho_{01}$. We note that taking the trace is not a strict requirement as it does not change the frequencies present in the Fourier transform and only slightly modifies the amplitudes, see Fig. S6.

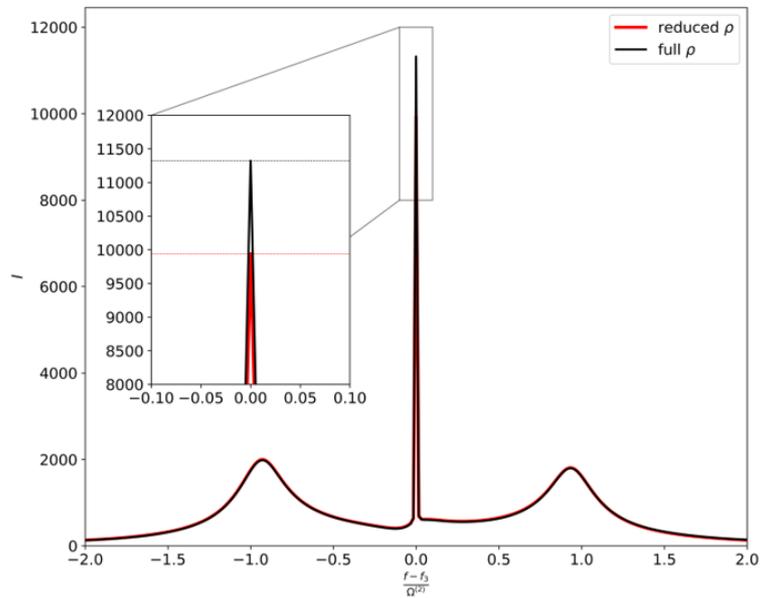

**Figure S6.** Frequency spectrum for a strongly driven ($V_{RF2}$ = 140 mV) spin obtained from the relevant off-diagonal element of the full (black) and reduced (red) density matrix. In both cases the splitting is clearly visible and identical, but the intensity of the center peak is slightly reduced in the Fourier transform of the reduced density matrix as shown in the inset.